\documentclass[
aps,%
12pt,%
final,%
notitlepage,%
oneside,%
onecolumn,%
nobibnotes,%
nofootinbib,%
showpacs,%
centertags]%
{revtex4}
\def\bb {\begin {eqnarray}}
\def\ee {\end {eqnarray}}

\usepackage[dvips]{epsfig,color}
\usepackage{graphicx}
\usepackage{amsmath,amsfonts,amsbsy,amssymb}
\usepackage{tabularx}

\begin{document}%\selectlanguage{english}
\title
{Supersymmetric Quantum Hall Liquid \\with a Deformed Supersymmetry
}

\author{Kazuki Hasebe}

\affiliation{Takuma National College of Technology}

\begin{abstract}

We construct a supersymmetric quantum Hall liquid with a deformed supersymmetry. 
One parameter is introduced in the supersymmetric Laughlin wavefunction to realize the original Laughlin wavefunction and the Moore-Read wavefunction in two extremal limits of the parameter. 
The introduced parameter corresponds to the coherence factor in the BCS theory. It is pointed out that the parameter-dependent supersymmetric Laughlin wavefunction enjoys a deformed supersymmetry. Based on the deformed supersymmetry, we construct a pseudo-potential Hamiltonian whose groundstate is exactly the parameter-dependent supersymmetric Laughlin wavefunction. 
Though the SUSY pseudo-potential Hamiltonian is parameter-dependent and non-Hermitian, its eigenvalues are parameter-independent and real. 

\end{abstract}


\maketitle

\section{Introduction}\label{int1}

The quantum Hall effects (QHE) are remarkable phenomena whose physical properties are deeply related to modern concepts of mathematics. 
The followings are some of the well-known examples.
The strong magnetic fields on QH systems bring a physical realization of non-commutative geometry \cite{GirvinPRB1984,arXiv:hep-th/0209198}. 
The edge states are chiral massless excitations, and are   
 described by the conformal field theory formalism \cite{WenPRL1990, StonePRB1990}. 
Excitations are anyons whose statistics is classified by the braid group \cite{WuPRL1984}. The effective field theory of QHE is given by the Chern-Simons topological field theory \cite{ZhangPRL1988,IJMP92B6}. 

In experiments, QH systems are realized in two-dimensional quantum wells. However, as mentioned above, since the QHE contains the deep and rich mathematical structures, one may tend to ask how the QHE and the related such mathematical structures will be generalized in higher dimensions.
In the past decade, there has been a great progress in the study of such   generalizations. 
The breakthrough was brought by Zhang and Hu's four-dimensional spherical set-up of the QHE \cite{cond-mat/0110572}.  
Their construction is based on the 2nd Hopf map, and indicated a first reasonable step toward generalizations of Haldane's QHE on a two-sphere \cite{PRL511983}. Their set-up was quickly extended on  more general higher dimensional manifolds \cite{hep-th/0606161}, such as complex projective spaces \cite{hep-th/0203264}, higher dimensional spheres \cite{cond-mat/0306045,hep-th/0310274}, and non-compact manifolds \cite{arXiv:hep-th/0505095}.

Recently, another direction of generalizations of QHE has also attracted attention: the supersymmetric (SUSY) extensions. 
Studies of one-particle problem in magnetic field on supermanifolds, $i.e.$ the SUSY Landau models, were launched by Ivanov et al.  
\cite{
hep-th/0311159,hep-th/0404108,hep-th/0510019,hep-th/0612300,arXiv:0705.2249,arXiv:0806.4716}. Independently, Hasebe and Kimura investigated Landau problem on supermanifolds \cite{hep-th/0409230,hep-th/0503162,arXiv:0809.4885}.
%\footnote{There are two different definitions of supersphere, one of which %is the coset $SU(2|1)/U(1|1)$ as used in \cite{hep-th/0311159} while the %other is $UOSp(1|2)/U(1)$ in \cite{hep-th/0409230}. 
% In this paper, we adopt the latter definition.}.
In such developments, particular properties of the SUSY Landau models are starting to be unveiled; non-anticommutative geometry in lowest Landau level  \cite{hep-th/0311159,hep-th/0404108,hep-th/0409230,hep-th/0503162,hep-th/0510019,arXiv:0809.4885}, 
enhanced SUSY in higher Landau levels 
\cite{hep-th/0503162,hep-th/0510019,hep-th/0612300,arXiv:0705.2249,arXiv:0806.4716}, existence of negative norm states and the remedy for it \cite{hep-th/0503162,hep-th/0510019,hep-th/0612300,arXiv:0705.2249,arXiv:0806.4716}.
Many-body problem on supermanifolds under magnetic fields, $i.e.$
 SUSY QHE, have also been explored \cite{hep-th/0503162,hep-th/0411137,arXiv:0705.4527,hep-th/0606007,arXiv:0710.0216,arXiv:0809.4885}. 
In Refs.\cite{hep-th/0411137,arXiv:0809.4885}, the SUSY Laughlin-Haldane wavefunctions are constructed so as to be invariant under given super Lie group symmetries.  
In this report, we introduce one-parameter family of the SUSY Laughlin-Haldane wavefunction.
The parameter-dependent SUSY Laughlin-Haldane wavefunction naturally reproduces the original Laughlin-Haldane wavefunction and the Moore-Read Pfaffian wavefunction in two extremal limits of the parameter.
Though the parameter-dependent Laughlin-Haldane wavefunction is not invariant under the original SUSY transformation, we show that 
there exists a deformed SUSY transformation under which the parameter-dependent SUSY  Laughlin-Haldane wavefunction is invariant. 
Based on the deformed supersymmetry, we construct a  
pseudo-potential Hamiltonian whose groundstate is exactly the parameter-dependent SUSY Laughlin-Haldane wavefunction. 

The paper is organized as follows. 
As a preliminary, we briefly explain the original set-up of the QHE on a  two-sphere in Sec.\ref{sec1}. In Sec.\ref{sec2}, we introduce the SUSY Hopf map and $UOSp(1|2)$ algebra.  
In Sec.\ref{sec3}, based on the SUSY Hopf map, we construct the SUSY Laughlin-Haldane wavefunction and its pseudo-potential Hamiltonian.  In Sec.\ref{sec4}, another but similar type of SUSY Laughlin-Haldane wavefunction is introduced based on a ``different''  SUSY Hopf map. In Sec.\ref{sec5}, a parameter-dependent deformed SUSY Hopf map and $UOSp(1|2) $ algebra are explored.
The parameter-dependent SUSY Laughlin-Haldane wavefunction and its pseudo-potential Hamiltonian are constructed in Sec.\ref{sec6}, and analogies to BCS state are discussed. Sec.\ref{sec7} is devoted to conclusions.

\section{The original bosonic Quantum Hall Liquid}\label{sec1}

Here, we briefly explain the QHE on a two-sphere formulated by Haldane \cite{PRL511983}.\footnote{With a non-compact version of the Hopf map, the QHE on a hyperboloid can also be developed 
\cite{arXiv:hep-th/0505095,arXiv:0809.4885}.}
The (1st) Hopf map is given by 
%%%%%%%%%%%%%%%%%%%%
\begin{equation}
S^3 \rightarrow S^2\simeq S^3/S^1,
\end{equation}
%%%%%%%%%%%%%%%%%%%%%%
where $S^n$ denotes $n$-dimensional sphere.
Explicitly, the Hopf map is realized as 
%%%%%%%%%%%%%%%%%%%%%%
\begin{equation}
\phi\rightarrow x_a=\phi^{\dagger} \sigma_a\phi,
\label{explicithopf}
\end{equation}
%%%%%%%%%%%%%%%%%%%%%
where $\sigma_a$ ($a=1,2,3$) are Pauli matrices, and $\phi$ is a two-component complex (Hopf) spinor which satisfies the normalization condition 
%%%%%%%%%%%%%%%%%%%%%%%
\begin{equation}
\phi^{\dagger}\phi=1.
\label{normalizationphi}
\end{equation}
%%%%%%%%%%%%%%%%%%%%%%%
With the constraint (\ref{normalizationphi}), $x_a$ satisfy the condition of $S^2$ with unit radius:   
%%%%%%%%%%%%%%%%%%%
\begin{equation}
x_ax_a=1.
\end{equation}
%%%%%%%%%%%%%%%%%%%%%
With the use of polar coordinates, the coordinates on a two-sphere are parameterized as 
%%%%%%%%%%%%%%%%%%%%%%
\begin{equation}
(x_1,x_2,x_3)=(\cos \varphi \sin\theta, \sin\varphi\sin\theta, \cos\theta),
\end{equation}
%%%%%%%%%%%%%%%%%%%%%%%
and the Hopf spinor is represented as 
%%%%%%%%%%%%%%%%%%%%%%%%%%%%%%%%%%%%%
\begin{equation}
\phi=\frac{1}{\sqrt{1+x_3}}
\begin{pmatrix}
1+x_3 \\
x_1+ix_2 
\end{pmatrix}e^{i\chi}=
\begin{pmatrix}
\cos \frac{\theta}{2} \\
\sin \frac{\theta}{2} e^{i\varphi}
\end{pmatrix}e^{i\chi},
\end{equation}
%%%%%%%%%%%%%%%%%%%%%%%%%
where the $U(1)$ phase factor $e^{i\chi}$ geometrically denotes $S^1$-fibre, and is canceled in the Hopf map (\ref{explicithopf}).  
The complex representation of the $SU(2)$ matrices $\tilde{\sigma}_a\equiv -\sigma_a^*$ is related to the original representation as 
%%%%%%%%%%%%%%%
\begin{equation}
\tilde{\sigma}_a =\epsilon\sigma_a \epsilon^{\dagger}
\end{equation}
%%%%%%%%%%%%%%%%
where $\epsilon$ is the two-rank antisymmetric tensor  
%%%%%%%%%%%%%%
\begin{equation}
\epsilon=i\sigma_2.
\end{equation}
%%%%%%%%%%%%%% 
The Laughlin-Haldane wavefunction is made from product of $SU(2)$ singlets of two Hopf spinors: 
%%%%%%%%%%%%%%%%%%%
\begin{equation}
\Phi=\prod_{i<j}(\phi_i \epsilon \phi_j)^m=
\prod_{i<j}(u_i v_j-v_i u_j)^m,
\label{bosonicLlinHaldanewave}
\end{equation}
%%%%%%%%%%%%%%%%%%%%
which is indeed invariant under the $SU(2)$ transformations generated by 
%%%%%%%%%%%%%%%
\begin{equation}
L_a=\frac{1}{2}\phi^t \tilde{\sigma}_a\frac{\partial}{\partial\phi}.
\end{equation}
%%%%%%%%%%%%%
The angular momentum of one-particle in the Laughlin-Haldane wavefunction is
 given by 
%%%%%%%%%%%%%%%%%%%%%%% 
\begin{equation}
L=\frac{1}{2}m(N-1).
\end{equation} 
%%%%%%%%%%%%%%%%%%%%
Similarly, the maximum angular momentum of any two-particles in the Laughlin-Haldane wavefunction is 
%%%%%%%%%%%%%%%%%%
\begin{equation}
J_{\text{max}}=2L-m=m(N-2).
\end{equation}
%%%%%%%%%%%%%%%%% 
%since any of the $SU(2)$ singlets in the Laughlin-Haldane wavefunction has %the power $m$. 
Namely,  the Laughlin-Haldane wavefunction does not contain two-body angular momentum that  exceeds $J_{\text{max}}$. 
Then, a pseudo-potential Hamiltonian whose zero energy groundstate is exactly the Laughlin-Haldane wavefunction is constructed as 
%%%%%%%%%%%%%%%%%%%%%%
\begin{equation}
H=\sum_{i<j}\sum_{J= J_{\text{max}}+1}^{2L}V_J P_J(i,j)
\end{equation}
%%%%%%%%%%%%%%%%%%%%%%%%
where $V_J$ is a positive coefficient, and $P_J$ denotes the projection operator to a subspace of two-body angular momentum $J$  
%%%%%%%%%%%%%%%%%
\begin{equation}
P_J(i,j)=\prod_{J'\neq J}\frac{C(i,j)-J'(J'+\frac{1}{2})}{J(J+{1})-J'(J'+{1})}.
\end{equation}
%%%%%%%%%%%%%%%%%%
Here, $C(i,j)$ is the $SU(2)$ Casimir operator  
%%%%%%%%%%%%%%%%%%%%%%
\begin{equation}
C(i,j)= (L_a(i)+L_a(j)) (L_a(i)+L_a(j))= 
2L_a(i) L_a(j)+2{L}({L}+1).
\end{equation}
%%%%%%%%%%%%%%%%%%%%%%

%%%%%%%%%%%%%%%%%%%%%%%%%%%%%%%%%%%%
%%%%%%%%%%%%%%%%%%%%%%%%%%%%%%%%%%%%
\section{The SUSY Hopf map and $UOSp(1|2)$ algebra}\label{sec2}
%%%%%%%%%%%%%%%%%%%%%%%%%%%%%%%%%%%%%%
%%%%%%%%%%%%%%%%%%%%%%%%%%%%%%%%%%%%%%

The SUSY extension of the Hopf map \cite{PLB193p61,J.Math.Phys.31(1990)} is given by\footnote{The non-compact versions of the SUSY Hopf map and the SUSY QHE were explored in Ref.\cite{arXiv:0809.4885}.}   
%%%%%%%%%%%%%
\begin{equation}
S^{3|2}\rightarrow S^{2|2}\simeq S^{3|2}/S^1. 
\end{equation}
%%%%%%%%%%%%%%%%%%%%
Explicitly, the SUSY Hopf map is realized as 
%%%%%%%%%%%%%%%%%%%%%%
\begin{equation}
\psi~~~\rightarrow ~~~x_a=\psi^{\ddagger}l_a\psi,~~~\theta^{\alpha}=\psi^{\ddagger}l_{\alpha}\psi, 
\end{equation}
%%%%%%%%%%%%%%%%%%%%%
where  $\psi$ is a three-component (SUSY Hopf) spinor, $\psi=(u,v,\eta)^t$, in which the first two-components are Grassmann even and the last component is Grassmann odd, and a  normalization condition is imposed as  
%%%%%%%%%%%%%%%%%%%%%%%
\begin{equation}
\psi^{\ddagger}\psi=1,
\label{normalizationofpsi}
\end{equation}
%%%%%%%%%%%%%%%%%%%%%%%
with $\psi^{\ddagger}=(u^*,v^*,-\eta^*)$\footnote{Here, $*$ denotes the superstar-conjugation, which acts to Grassmann-odd quantities as $(\eta^*)^*=-\eta$ and $(\eta_1\eta_2)^*=\eta_1^*\eta_2^*$.}. 
$l_a$ and $l_{\alpha}$ are   
%%%%%%%%%%%%%%%%%%%%%%%%%%%%
\begin{equation}
l_a=\frac{1}{2}\begin{pmatrix}
\sigma_a & 0 \\
0 & 0 
\end{pmatrix},
~~l_{\alpha}=\frac{1}{2}
\begin{pmatrix}
0 & \tau_{\alpha} \\
-(\epsilon \tau_{\alpha})^t & 0 
\end{pmatrix},
\end{equation}
%%%%%%%%%%%%%%%%%%%%%%%%%%%%
with $\tau_1=(1,0)$ and $\tau_2=(0,1)^t$, and  
they satisfy the $UOSp(1|2)$ algebra\footnote{
The $UOSp(1|2)$ Casimir operator is given by
%%%%%%%%%%%%%%%%%%%%%
\begin{equation}
C=L_a L_a+\epsilon_{\alpha\beta}L_{\alpha}L_{\beta}.  
\end{equation}
%%%%%%%%%%%%%%%%%%%
It is noticed that the fermionic generators $L_{\alpha}$ are not Hermitian (but pseudo-Hermitian), and the Casimir operator is not either. However, the eigenvalues  of $UOSp(1|2)$ Casimir operator are real and explicitly given by $L(L+\frac{1}{2})$ with $L=0,{1}/{2},1,{3}/{2},\cdots$. 
The irreducible decomposition rule of $UOSp(1|2)$ is  
%%%%%%%%%%%%%%%%%%%%%%%
\begin{equation}
L\otimes L=0\oplus \frac{1}{2} \oplus 1 \oplus \frac{3}{2} \oplus  \cdots \oplus 2L-\frac{1}{2} \oplus 2L.
\end{equation}
%%%%%%%%%%%%%%%%%%%%%%%
}
%%%%%%%%%%%%%%%%
\begin{equation}
[l_a,l_b]=i\epsilon_{abc}l_c,~~[l_a,l_{\alpha}]=\frac{1}{2}(\sigma_a)_{\beta\alpha}l_{\beta},~~\{l_{\alpha},l_{\beta}\}=\frac{1}{2}(\epsilon \sigma_a)_{\alpha\beta}l_a.
\label{oneSUSYalgebra}
\end{equation}
%%%%%%%%%%%%%%%%%%%
 $l_{\alpha}$ satisfy the pseudo-Hermitian condition\footnote{ The superadjoint $\ddagger$ is defined as 
%%%%%%%%%%%%%%%%%%
\begin{equation}
\begin{pmatrix}
A & B \\
C & D
\end{pmatrix}^{\ddagger}
=
\begin{pmatrix}
A^{\dagger} & C^{\dagger} \\
-B^{\dagger} & D^{\dagger}
\end{pmatrix},
\end{equation}
%%%%%%%%%%%%%%%%%%%
where $A$ and $D$ are Grassmann-even component matrices, while $B$ and $C$ are Grassmann-odd component matrices. }
%%%%%%%%%%%%%%%%%%%%%%%
\begin{equation}
{l_{\alpha}}^{\ddagger}=\epsilon_{\alpha\beta}l_{\beta},
\end{equation}
%%%%%%%%%%%%%%%%%%%%%%
and $\theta_{\alpha}$ satisfy the pseudo-real condition $\theta_{\alpha}^{*}=\epsilon_{\alpha\beta}\theta_{\beta}$. 
With the constraint (\ref{normalizationofpsi}), it is a simple task to 
show that $x_a$ and $\theta_{\alpha}$ satisfy the condition 
%%%%%%%%%%%%%%%%%%%
\begin{equation}
x_ax_a+\epsilon_{\alpha\beta}\theta_{\alpha}\theta_{\beta}=1,
\end{equation}
%%%%%%%%%%%%%%%%%%%%%
which defines $S^{2|2}$ with unit radius. 
The explicit form of the SUSY Hopf spinor is derived as  
%%%%%%%%%%%%%%%%%%%%%%
\begin{equation}
\psi=\frac{1}{\sqrt{2(1+x_3)}}
\begin{pmatrix}
(1+x_3) (1-\frac{1}{4(1+x_3)}\theta\epsilon\theta) \\
(x_1+ix_2) (1+\frac{1}{4(1+x_3)}\theta\epsilon\theta)    \\
(1+x_3)\theta_1 +(x_1+ix_2)\theta_2
\end{pmatrix} e^{i\chi}.
\end{equation}
%%%%%%%%%%%%%%%%%%%%%%%%%
The complex representation of $UOSp(1|2)$ generators are given by 
%%%%%%%%%%%%%%%%%%%%
\begin{equation}
\tilde{l}_a=-{l_a}^{*},~~\tilde{l}_{\alpha}=\epsilon_{\alpha\beta}l_{\beta},
\end{equation}
%%%%%%%%%%%%%%%%%%%%
and they are related to the original representation as 
%%%%%%%%%%%%%%%%%%%%%%%
\begin{equation}
\tilde{l}_a=\mathcal{R}l_a \mathcal{R}^{\dagger},~~\tilde{l}_{\alpha}=\mathcal{R}l_{\alpha}\mathcal{R}^{\dagger },
\end{equation}
%%%%%%%%%%%%%%%%%%%%%%%%%%
with  
%%%%%%%%%%%%%%%%%%%%%
\begin{equation}
\mathcal{R}=
\begin{pmatrix}
0 & 1 & 0 \\
-1 & 0 & 0 \\
0 & 0 & -1
\end{pmatrix}.
\end{equation}
%%%%%%%%%%%%%%%%%%%%%%%

%%%%%%%%%%%%%%%%%%%%%%%%%%%%%%%%%
%%%%%%%%%%%%%%%%%%%%%%%%%%%%%%%%%
\section{SUSY Quantum Hall Liquid}\label{sec3}
%%%%%%%%%%%%%%%%%%%%%%%%%%%%%%%%%%
%%%%%%%%%%%%%%%%%%%%%%%%%%%%%%%%%%%

As the original Laughlin-Haldane wavefunction is invariant under the $SU(2)$ transformation, a SUSY extension of the Laughlin-Haldane wavefunction is constructed so as to be invariant under the $UOSp(1|2)$ transformations \cite{hep-th/0411137}: 
%%%%%%%%%%%%%%%%%%%
\begin{equation}
\Psi=\prod_{i<j}(\psi_i \mathcal{R}\psi_j)^m=
\prod_{i<j}(u_i v_j-v_i u_j-\eta_i\eta_j)^m.
\label{theoriginalLlin}
\end{equation}
%%%%%%%%%%%%%%%%%%%% 
 $\Psi$ is indeed invariant under the $UOSp(1|2)$ transformations generated by 
%%%%%%%%%%%%%%%%%%%%%%%%%%%%
\begin{equation}
L_a=\psi^t\tilde{l}_a\frac{\partial}{\partial\psi},~~
L_{\alpha}=\psi^t \tilde{l}_{\alpha}\frac{\partial}{\partial\psi}.
\label{generatorseffectiveLLL}
\end{equation}
%%%%%%%%%%%%%%%%%%%%%%%%%%
The pseudo-potential Hamiltonian whose exact groundstate is the SUSY  Laughlin-Haldane wavefunction can be similarly constructed by  the similar procedure discussed in Sec.\ref{sec1}. 
The maximum of the two-particle $UOSp(1|2)$ angular momentum in the SUSY Laughlin-Haldane wavefunction is $J_{\text{max}}=m(N-2)$, and the pseudo-potential Hamiltonian is given by  
%%%%%%%%%%%%%%%%%%%%%%
\begin{equation}
H=\sum_{i<j}\sum_{J=J_{\text{max}}+\frac{1}{2}}^{2L}V_J P_J(i,j),
\label{pseudopotham}
\end{equation}
%%%%%%%%%%%%%%%%%%%%%%%%
where the projection operator 
%%%%%%%%%%%%%%%%%
\begin{equation}
P_J(i,j)=\prod_{J'\neq J}\frac{C(i,j)-J'(J'+\frac{1}{2})}{J(J+\frac{1}{2})-J'(J'+\frac{1}{2})}
\label{ospprojectionop}
\end{equation}
%%%%%%%%%%%%%%%%%%
is constructed by the $UOSp(1|2)$ Casimir operator  
%%%%%%%%%%%%%%%%%%%%%%
\begin{align}
&C(i,j)= (L_a(i)+L_a(j)) (L_a(i)+L_{a}(j))+\epsilon_{\alpha\beta}(L_{\alpha}(i)+L_{\alpha}(j))(L_{\beta}(i)+L_{\beta}(j))\nonumber\\
&~~~~~~~~~= 
2L_a(i) L_a(j)+2\epsilon_{\alpha\beta}L_{\alpha}(i)L_{\beta}(j)+2L(L+\frac{1}{2}).
\end{align}
%%%%%%%%%%%%%%%%%%%%%%
It should be noticed that the pseudo-potential Hamiltonian is not Hermitian, since the fermionic generators $L_{\alpha}$ are (pseudo-Hermitian but) not Hermitian. However, the eigenvalues of the pseudo-Hamiltonian are real.

%%%%%%%%%%%%%%%%%%%%%%%%%%%%%%%%%%%%%%%%%%%%%%%%%%%%%%%%%%%%%%%%%%
%%%%%%%%%%%%%%%%%%%%%%%%%%%%%%%%%%%%%%%%%%%%%%%%%%%%%%%%%%%%%%%%%%
\section{Another SUSY Quantum Hall Liquid}\label{sec4}
%%%%%%%%%%%%%%%%%%%%%%%%%%%%%%%%%%%%%%%%%%%%%%%%%%%%%%%%%%%%%%%%%%
%%%%%%%%%%%%%%%%%%%%%%%%%%%%%%%%%%%%%%%%%%%%%%%%%%%%%%%%%%%%%%%%%% 

With 
%%%%%%%%%%%%%%%%%%%%%%
\begin{equation}
d_{\alpha}=-\frac{1}{2}
\begin{pmatrix}
0 & \tau_{\alpha}\\
(\epsilon \tau_{\alpha})^t & 0
\end{pmatrix},~~~\gamma=
\begin{pmatrix}
1 & 0 & 0 \\
0 & 1  & 0 \\
0 & 0 & 2
\end{pmatrix},
\end{equation}
%%%%%%%%%%%%%%%%%%%%%%%%
 $l_{a}$ and $l_{\alpha}$ satisfy the following $SU(2|1)$ algebra: 
%%%%%%%%%%%%%%%%%%%%%%%%
\begin{align}
&[l_a,d_{\alpha}]=\frac{1}{2}(\sigma_a)_{\beta\alpha}d_{\beta},~~~~\{d_{\alpha},d_{\beta}\}=-\frac{1}{2}(\epsilon \sigma_a)_{\alpha\beta}
l_a,~~~\{l_{\alpha},d_{\beta}\}=-\frac{1}{4}\epsilon_{\alpha\beta}\gamma,\nonumber\\
&[\gamma,l_a]=0,~~~~[\gamma,l_{\alpha}]=-d_{\alpha},
~~~~[\gamma,d_{\alpha}]=-l_{\alpha}.
\end{align}
%%%%%%%%%%%%%%%%%%%%% 
$d_{\alpha}$ satisfy the pseudo-Hermitian conjugation relation 
%%%%%%%%%%%%%%%%%%%%%%%
\begin{equation}
{d_{\alpha}}^{\ddagger}=-\epsilon_{\alpha\beta}d_{\beta}.
\end{equation}
%%%%%%%%%%%%%%%%%%%%%%
$l_{\alpha}$ and $d_{\alpha}$ are linearly independent, but they are related by the Hermitian conjugation 
%%%%%%%%%%%%%%%%%%%
\begin{equation}
{l_{\theta_1}}^{\dagger}=-d_{\theta_2},~~~{l_{\theta_2}}^{\dagger}=d_{\theta_1}.
\end{equation}
%%%%%%%%%%%%%%%%%%%%%%

One may find that $l_a$ and $d_{\alpha}$ satisfy the closed algebra: 
%%%%%%%%%%%%%%%%%%%%%%%%%%
\begin{equation}
[l_a,l_b]=i\epsilon_{abc}l_c, 
~~
[l_a,d_{\alpha}]=\frac{1}{2}(\sigma_a)_{\beta\alpha}d_{\beta},
~~
\{d_{\alpha},d_{\beta}\}=-\frac{1}{2}(\epsilon \sigma_a)_{\alpha\beta}l_a.
\end{equation}
%%%%%%%%%%%%%%%%%%%%%%%%%%%
The complex representation $\tilde{d}_{\alpha}=\epsilon_{\alpha\beta}d_{\beta}$ is related to the original representation as 
%%%%%%%%%%%%%%%%%
\begin{equation}
\tilde{d}_{\alpha}=\mathcal{R}d_{\alpha}\mathcal{R}^{\dagger}. 
\end{equation}
%%%%%%%%%%%%%%%%%
With use of $l_a$ and $d_{\alpha}$, another SUSY Hopf map may be constructed  
as 
%%%%%%%%%%%%%%%%%%%%%%
\begin{equation}
\psi \rightarrow x_a =2\psi^{\ddagger}l_a\psi,~~~\theta'_{\alpha}=2\psi^{\ddagger}d_{\alpha}\psi,
\end{equation}
%%%%%%%%%%%%%%%%%%%%
where $\psi$ is subject to th the normalization condition (\ref{normalizationofpsi}). It is readily shown that $x_a$ and $\theta'_{\alpha}$ satisfy the condition $x_a x_a -\epsilon_{\alpha\beta}\theta'_{\alpha}\theta'_{\beta}=1$. 

Another SUSY Laughlin-Haldane wavefunction that is invariant under the transformations generated by $L_a=\psi^t \tilde{l}_a \frac{\partial}{\partial\psi}$ and $D_{\alpha}=\psi^t \tilde{d}_{\alpha}\frac{\partial}{\partial\psi} $ is derived as  
%%%%%%%%%%%%%%%%%%%%%%
\begin{equation}
\Psi'=\prod_{i<j}(-\psi^t_i \mathcal{R}^t\psi_j)^m=
\prod_{i<j}(u_i v_j-v_i u_j+\eta_i \eta_j)^m.
\label{anotherLlin}
\end{equation}
%%%%%%%%%%%%%%%%%%%%%%
$\Psi'$ is different from $\Psi$ by the sign in front of the fermion bilinear term. 
The corresponding pseudo-potential Hamiltonian and the projection operator  can be similarly constructed,  
 and  take the same forms of (\ref{pseudopotham}) and 
(\ref{ospprojectionop}), respectively. But, in the present, the Casimir operator is made from $L_a$ and $D_{\alpha}$ as 
%%%%%%%%%%%%%%%%%%%%%%
\begin{align}
&C'(i,j)= (L_a(i)+L_a(j)) (L_a(i)+L_a(j))-\epsilon_{\alpha\beta}(D_{\alpha}(i)+D_{\alpha}(j))(D_{\beta}(i)+D_{\beta}(j))\nonumber\\
&~~~~~~~~~= 
2L_a(i) L_a(j)-2\epsilon_{\alpha\beta}D_{\alpha}(i)D_{\beta}(j)+2L({L}+\frac{1}{2}).
\end{align}
%%%%%%%%%%%%%%%%%%%%%%

%%%%%%%%%%%%%%%%%%%%%%%%%%%%%%%%%%%%%%%%%%%%%%%%%%%%%%%
\section{Deformed SUSY Hopf map and Supersymmetry}\label{sec5}
%%%%%%%%%%%%%%%%%%%%%%%%%%%%%%%%%%%%%%%%%%%%%%%%%%%%%%%%

In Sec.\ref{sec3} and Sec.\ref{sec4}, we encountered two ``different'' SUSY Laughlin-Haldane wavefunctions. One may speculate that there will be a general SUSY wavefunctions from which two such SUSY Laughlin-Haldane wavefunctions are naturally reproduced. 
For the construction of such a general SUSY wavefunction, we first introduce a parameter-dependent  SUSY Hopf map  
%%%%%%%%%%%%%%%%%%%%%%%%%%%%
\begin{equation}
\psi \rightarrow x_a= \psi^{\ddagger} l_a\psi,~~~\theta_{\alpha}(x) =\psi^{\ddagger}v_{\alpha}(x)\psi, 
\label{deformedSUSYHopfmap}
\end{equation}
%%%%%%%%%%%%%%%%%%%%%%%%%%%%%
where 
%%%%%%%%%%%%%%%%%%%%%%%%
\begin{eqnarray}
l_a=\frac{1}{2}
\begin{pmatrix}
\sigma_a & 0 \\
0 & 0 
\end{pmatrix},~~~
v_{\alpha}(x)=
\frac{1}{2}
\begin{pmatrix}
0 & {x}\tau_{\alpha}\\
-\frac{1}{x} (\epsilon \tau_{\alpha})^t & 0
\end{pmatrix}.
\end{eqnarray}
%%%%%%%%%%%%%%%%%%%%%%%%
$v(x)$ depends on a complex parameter $x$ ($x$ has nothing to do with the bosonic coordinates $x^a$), and $\theta_{\alpha}(x)$ as well. Though $\theta_{\alpha}(x)$ depends on the parameter $x$, $x_a$ and $\theta_{\alpha}(x)$ superficially satisfy the parameter-$\it{independent}$  supersphere condition:  
%%%%%%%%%%%%%%
\begin{equation}
x_a x_a +\epsilon_{\alpha\beta}\theta_{\alpha}(x)\theta_{\beta}(x)=1.
\end{equation}
%%%%%%%%%%%%%
However, $\theta_{\alpha}(x)$ should not generally be regarded as fermionic coordinates on a supersphere, since $\theta_{\alpha}(x)$ are not pseudo-real unless $|x|=1$. 
$v_{\alpha}(x)$ are constructed by a linear combination of $l_{\alpha}$ and $d_{\alpha}$;  
%%%%%%%%%%%%%%%%%%%%%%
\begin{equation}
v_{\alpha}(x)= \frac{1}{2x}(l_{\alpha}+d_{\alpha})+\frac{x}{2}(l_{\alpha}-d_{\alpha}) =
\biggl(\frac{x}{2}+\frac{1}{2x}\biggr)l_{\alpha}-\biggl(\frac{x}{2}-\frac{1}{2x}\biggr)d_{\alpha}.
\end{equation}
%%%%%%%%%%%%%%%%%%%%%%
$v_{\alpha}(x)$ is reduced to $l_{\alpha}$ at $x=1$, and  $id_{\alpha}$ at $x=-i$. $v_{\alpha}(x)$ is not well defined in the limits $x\rightarrow 0$ and $x\rightarrow \infty$, because the matrix elements of $v_{\alpha}$ diverge. (These two extremal limits correspond to the original Laughlin state and the Moore-Read state as we shall see in Sec.\ref{sec6}.)
In general, $v_{\alpha}(x)$ are not pseudo-Hermitian since 
%%%%%%%%%%%%%%%%%%%
\begin{equation}
{v_{\alpha}(x)}^{\ddagger}=\epsilon_{\alpha\beta}v_{\beta}({1}/{x^*}).
\end{equation}
%%%%%%%%%%%%%%%%%%%
$v_{\alpha}(x)$ becomes  pseudo-Hermitian $v_{\alpha}(x)^{\ddagger}=\epsilon_{\alpha\beta}v_{\beta}(x)$, if and only if  
$x$ satisfies $|x|=1$. When we parameterize $x=e^{i\omega}$, $v_{\alpha}$ can be expressed as  
$v_{\alpha}(e^{i\omega})=l_{\alpha}\cos \omega  -id_{\alpha}\sin \omega.$ 

$l_a$ and $v_{\alpha}(x)$ are related to the original $UOSp(1|2)$ generators as 
%%%%%%%%%%%%%%%%%%%%%%%%%%
\begin{equation}
l_a=g(1/x) l_{\alpha}g(x),~~~v^{\alpha}(x)=g(1/x)l^{\alpha}g(x)
\end{equation}
%%%%%%%%%%%%%%%%%%%%%%%%%%%
with 
%%%%%%%%%%%%%%%%%%%
\begin{equation}
g(x)=
\begin{pmatrix}
1 & 0 & 0 \\
0 & 1 & 0 \\
0 & 0 & x
\end{pmatrix},
~~~g(1/x)=  
\begin{pmatrix}
1 & 0 & 0 \\
0 & 1 & 0 \\
0 & 0 & 1/x
\end{pmatrix}
=g(x)^{-1}.
\end{equation}
%%%%%%%%%%%%%%%%%%%%
Though $v_{\alpha}(x)$ depends on the parameter $x$,  $l_a$ and $v_{\alpha}(x)$ satisfy the parameter-$\it{independent}$ $UOSp(1|2)$ algebraic relations
\footnote{In this sense, the present algebra is different from $q$-deformed algebra.
In $q$-deformed algebra, the parameter $q$ explicitly appears in the algebraic relations. Meanwhile, in the present, though the $UOSp(1|2)$  generators depend on the parameter $x$, they form parameter-independent algebraic relations.}:
%%%%%%%%%%%%%%%%%%%%%%%%%%%
\begin{eqnarray}
[l_a,l_b]=i\epsilon_{abc}l_c,~~
[l_a,v_{\alpha}(x)]=\frac{1}{2}(\sigma_a)_{\beta\alpha}v_{\beta}(x),~~
\{v_{\alpha}(x),v_{\beta}(x)\}=\frac{1}{2}(\epsilon \sigma_a)_{\alpha\beta}l_a.
\end{eqnarray}
%%%%%%%%%%%%%%%%%%%%%%%%%%%
Then,  $v_{\alpha}(x)$ are regarded as  ``ordinary'' fermionic generators of $UOSp(1|2)$, except for the pseudo-Hermiticity.  
The complex representation, $\tilde{l}_a\equiv -{l_a}^*$ and $\tilde{v}_{\alpha}(x)\equiv \epsilon_{\alpha\beta}v_{\beta}(x)$, is related to the original representation as 
%%%%%%%%%%%%%%%%
\begin{equation}
\tilde{l}_a=\mathcal{R}l_a \mathcal{R}^{\dagger},~~
\tilde{v}_{\alpha}(x)=\mathcal{R}v_{\alpha}(x) \mathcal{R}^{\dagger}.
\end{equation}
%%%%%%%%%%%%%%%% 
Since $v_{\alpha}$ depends on the parameter $x$, the corresponding Casimir operator 
%%%%%%%%%%%%%%%%%%%%%%%%%%%%%
\begin{equation}
C=l_al_a+\epsilon_{\alpha\beta}v_{\alpha}(x)v_{\beta}(x)
\end{equation}
%%%%%%%%%%%%%%%%%%%%%%%%%%%%%
also does the parameter. However,  the eigenvalues of the Casimir operator 
are parameter-$\it{independent}$ (and given by $L(L+1/2)$ with $L=0,1/2,1,3/2,\cdots$),  since  $l_a$ and $v_{\alpha}(x)$ satisfy the parameter-independent $UOSp(1|2)$ algebraic relations.

%%%%%%%%%%%%%%%%%%%%%%%%%%%%%%%%%%%%%%%%%%%%%%%%%%%%%%%%%%%%%%%%%%%%%
%%%%%%%%%%%%%%%%%%%%%%%%%%%%%%%%%%%%%%%%%%%%%%%%%%%%%%%%%%%%%%%%%%%%
\section{Deformed SUSY Quantum Hall Liquid}\label{sec6}
%%%%%%%%%%%%%%%%%%%%%%%%%%%%%%%%%%%%%%%%%%%%%%%%%%%%%%%%%%%%%%%%%%%%%
%%%%%%%%%%%%%%%%%%%%%%%%%%%%%%%%%%%%%%%%%%%%%%%%%%%%%%%%%%%%%%%%%%%%%

The deformed SUSY Hopf map (\ref{deformedSUSYHopfmap}) can be rewritten as 
%%%%%%%%%%%%%%%%%%%%%%%%%%%
\begin{equation}
\psi(x) \rightarrow x_a= \psi^{\ddagger}(1/x^*) l_{a}\psi(x) ,~~~\theta_{\alpha}(x) =\psi^{\ddagger}({1}/{x^*})l_{\alpha}\psi(x), 
\label{deformSUSYHopf}
\end{equation}
%%%%%%%%%%%%%%%%%%%%%%%%%%%%
where $\psi(x)$ and $\psi^{\ddagger}(x)$ are defined as 
%%%%%%%%%%%%%%%%%%%%%%%%%%%%%%%%%
\begin{equation}
\psi(x)=(u,v,x\eta)^t,~~~\psi^{\ddagger}(1/x^*)=(u^*,v^*,-\eta^*/x ), 
\end{equation}
%%%%%%%%%%%%%%%%%%%%%%%%%%%%%%%
and satisfy $\psi^{\ddagger}(1/x^*)\psi(x)=1$. (It should be noted, when $v^{\alpha}(x)$ is pseudo-real, $i.e.$ $|x|=1$, $x=e^{i\omega}$ can be absorbed by redefinition of $\eta$, and (\ref{deformSUSYHopf}) is reduced to the original SUSY Hopf map.)
With use of the parameter-dependent SUSY Hopf spinor $\psi(x)$, we introduce a parameter-dependent SUSY Laughlin-Haldane wavefunction 
%%%%%%%%%%%%%%%%%%%%
\begin{equation}
\Psi^{(r)}=\prod_{i<j}(\psi(x)_i^t \mathcal{R}\psi(x)_j)^m=\prod_{i<j}(u_iv_j-v_iu_j-r \eta_{i}\eta_{j})^m,
\label{parameterdependLaughlin}
\end{equation}
%%%%%%%%%%%%%%%%%%%%%
with 
%%%%%%%%%%%%%%%%%%
\begin{equation}
r=x^2.
\end{equation} 
%%%%%%%%%%%%%%%%%%%%
At $r=1$, $\Psi^{(r)}$ reproduces $\Psi$ (\ref{theoriginalLlin}), while at $r=-1$, $\Psi^{(r)}$ does $\Psi'$ (\ref{anotherLlin}).
It is straightforward to check that $\Psi^{(r)}$ is invariant under the deformed $UOSp(1|2)$ transformations generated by   
%%%%%%%%%%%%%%%%%%%%%%%%%%
\begin{equation}
L_{\alpha}=\psi^t(x)\tilde{l}_a\frac{\partial}{\partial\psi(x)} =\psi^t \tilde{l}_a \frac{\partial}{\partial\psi},~~~ V_{\alpha}=\psi^t(x)\tilde{l}_{\alpha}\frac{\partial}{\partial\psi(x)}=\psi^t \tilde{v}_{\alpha}({1}/{x})\frac{\partial}{\partial\psi}.
\end{equation}
%%%%%%%%%%%%%%%%%%%%%%%%%%%
Taking advantage of the deformed $UOSp(1|2)$ symmetry of $\Psi^{(r)}$, 
the pseudo-potential Hamiltonian for $\Psi^{(r)}$ can be constructed as   
%%%%%%%%%%%%%%%%%%%%%%
\begin{equation}
H^{(r)}=\sum_{i<j}\sum_{J\ge J_{\text{max}}+\frac{1}{2}}^{2L}V_J P^{(r)}_J(i,j),
\end{equation}
%%%%%%%%%%%%%%%%%%%%%%%%
where the projection operator is given by 
%%%%%%%%%%%%%%%%%
\begin{equation}
P^{(r)}_J(i,j)=\prod_{J'\neq J}\frac{C^{(r)}(i,j)-J'(J'+\frac{1}{2})}{J(J+\frac{1}{2})-J'(J'+\frac{1}{2})},
\end{equation}
%%%%%%%%%%%%%%%%%%
and the parameter-dependent Casimir operator is    
%%%%%%%%%%%%%%%%%%%
\begin{align}
&C^{(r)}(i,j)= (L_a(i)+L_a(j)) (L_a(i)+L_{a}(j))+\epsilon_{\alpha\beta}(V_{\alpha}(i)+V_{\alpha}(j))(V_{\beta}(i)+V_{\beta}(j))\nonumber\\
&~~~~~~~~~= 
2L_a(i) L_a(j)+2\epsilon_{\alpha\beta}V_{\alpha}(i)V_{\beta}(j)+2L(L+\frac{1}{2}).
\end{align}
%%%%%%%%%%%%%%%%%%%%%%
Since eigenvalues of the Casimir operator do not depend on the parameter, the eigenvalues of the pseudo-potential Hamiltonian do not either. 
%Further, the pseudo-potential Hamiltonian is $\it{not}$ Hermitian due to 
%the existence of the non-Hermitian fermionic operators $V_{\alpha}$, but  
%eigenvalues of pseudo-potential Hamiltonian are real.

Next, we discuss physical meaning of the parameter $r$. 
The parameter-dependent SUSY Laughlin-Haldane wavefunction can be  rewritten as 
%%%%%%%%%%%%%%%%%%%%%
\begin{equation}
\Psi^{(r)}=e^{ -mr\sum_{i<j}\frac{\eta^i\eta^j}{u_iv_j-v_i u_j}}\Phi,
\label{exponentSUSYLlin}
\end{equation}
%%%%%%%%%%%%%%%%%%%%%
where $\Phi$ is the original Laughlin wavefunction (\ref{bosonicLlinHaldanewave}). 
Expanding  the exponential in (\ref{exponentSUSYLlin}), we obtain 
%%%%%%%%%%%%%%%%%%%%%%%%
\begin{align}
&\Psi^{(r)}=\Phi -mr\sum_{i<j}\frac{\eta^i\eta^j}{u_iv_j-v_i u_j} \Phi+\frac{1}{2} (mr)^2(\sum_{i<j}\frac{\eta^i\eta^j}{u_iv_j-v_i u_j})^2 \Phi\nonumber\\
&~~~~~~+\cdots + (-mr)^{N/2} \eta_1\eta_2\cdots\eta_N \cdot Pf(\frac{1}{u_iv_j-v_iu_j})\Phi.
\label{expansionSUSYLlin}
\end{align}
%%%%%%%%%%%%%%%%%%%%%%%%
This expansion is formally regarded as a perturbative expansion about the parameter $r$. ($\Phi$ itself depends on the parameter $m$, and then, $m$ should not be taken as the expansion parameter.) 
%As mentioned above, the two extremal limits $r\rightarrow 0$ and $r%\rightarrow \infty$ are special in the sense that the 
%SUSY generators are not well defined. 
In the limit  $r\rightarrow 0$, the SUSY Laughlin-Haldane wavefunction is reduced to 
%%%%%%%%%%%%%%%%%
\begin{equation}
\Psi^{(r)} \rightarrow \Phi=\prod_{i<j}(u_iv_j-v_i u_j)^m, 
\end{equation}
%%%%%%%%%%%%%%%%%%
while in the limit $r\rightarrow \infty$,  
%%%%%%%%%%%%%%%%%
\begin{equation}
\Psi^{(r)} \rightarrow (-mr)^{N/2}  \eta_1\eta_2\cdots \eta_{N}\cdot  Pf(\frac{1}{u_iv_j-v_iu_j})\Phi.
\end{equation}
%%%%%%%%%%%%%%%%%
Here, $Pf$ represents the Pfaffian function, and  $Pf(\frac{1}{u_iv_j-v_iu_j})\Phi$ is known as the Moore-Read wavefunction, which describes the $p$-wave pairing groundstate of the QHE at $\nu=5/2$ for $m=2$ \cite{NPB360362,PRL663205}. 
Thus, in the two extremal limits of the SUSY Laughlin-Haldane wavefunction, two different QH groundstate wavefunctions are naturally realized. 
There also exist interesting analogies between the SUSY QHE and the BCS superconductivity \cite{arXiv:0705.4527}, and their analogies become clear by introducing the parameter $r$. 
The BCS state of superconductivity is given by 
%%%%%%%%%%%%%%%%%%%%%%%%%%
\begin{equation}
|\text{BCS}\rangle=\prod_k(1+g_k c_{k\uparrow}^{\dagger} c_{-k\downarrow}^{\dagger})|0\rangle,
\end{equation}
%%%%%%%%%%%%%%%%%%%%%%%%%%%%%
where $c^{\dagger}_{k\sigma}$ denotes an electron-creation operator with momentum $k$ and spin $\sigma$. 
The BCS state is rewritten as 
%%%%%%%%%%%%%%%%%%%%%%%%%
\begin{equation}
|\text{BCS}\rangle=e^{ \sum_k g_k c_{k\uparrow}^{\dagger}c_{-k\downarrow}^{\dagger} } |0\rangle,
\label{exponentBCS}
\end{equation}
%%%%%%%%%%%%%%%%%%%%%%
and expanded as 
%%%%%%%%%%%%%%%%%%%
\begin{equation}
|\text{BCS}\rangle=|0\rangle+ \sum_k g_k c_{k\uparrow}^{\dagger}c_{-k\downarrow}^{\dagger} |0\rangle+\frac{1}{2}(\sum_k g_k c_{k\uparrow}^{\dagger}c_{-k\downarrow}^{\dagger})^2|0\rangle+\cdots+\prod_{k}g_k c^{\dagger}_{k\uparrow}c^{\dagger}_{-k\downarrow}|0\rangle.
\label{expansionofBCS}
\end{equation}
%%%%%%%%%%%%%%%%%%%%%
One may find apparent analogies between (\ref{exponentSUSYLlin}) and (\ref{exponentBCS}), and between (\ref{expansionSUSYLlin}) and (\ref{expansionofBCS}). 
In the BCS state, the number of  Cooper pairs fluctuates with finite $g_k$, and the BCS state represents the superconducting state.
With two extremal limits $g_k=0$ and $g_k\rightarrow \infty$, the electron-number does not fluctuate, and  the BCS state is reduced to the no-electron vacuum and the filled Fermi sphere, respectively.  The analogies between the BCS state and the SUSY Laughlin-Haldane wavefunction are summarized in Table 
I.

%%%%%%%%%%%%%%%%%%%%%%%%%%%%%%%%%%%%%%%%%%%%%%%
%%%%%%%%%%%%%%%%%%%%%
\begin{table} 
\renewcommand{\arraystretch}{1}
\begin{center}
\begin{tabular}{|c|c|c|}  
 \hline    & BCS state &  SUSY Laughlin state  
\\
\hline \hline
paring  & $s$-wave &   $p$-wave
\\
\hline  
parameter & coherence factor $g_k$ &    $r$ 
\\
 \hline 
``Groundstate''  &  no-electron vacuum &   Laughlin state    
\\ 
\hline    
  ``Filled state'' &  Filled Fermi sphere & Moore-Read state \\ 
\hline
\end{tabular} 
\end{center}
\caption{Analogies between the BCS state  and the SUSY Laughlin-Haldane wavefunction.} 
\end{table} 
%%%%%%%%%%%%%%%%%%%%%%%%%%%%%%%%%%%%%%%%%%%%%%%%%%%%%%%%%%%%%%%%%%%%%%

%%%%%%%%%%%%%%%%%%%%%%%%%%%%%%%%%%%%%%
%%%%%%%%%%%%%%%%%%%%%%%%%%%%%%%%%%%%%%
\section{Conclusions}\label{sec7}
%%%%%%%%%%%%%%%%%%%%%%%%%%%%%%%%%%%%%%
%%%%%%%%%%%%%%%%%%%%%%%%%%%%%%%%%%%%%%%%

Developing a parameter-dependent SUSY Hopf map, we constructed a parameter-dependent SUSY Laughlin-Haldane wavefunction.
I have shown that the parameter-dependent Laughlin-Haldane wavefunction respects a deformed supersymmetry. Based on the deformed supersymmetry, we derived the corresponding SUSY pseudo-potential Hamiltonian. 
The parameter represents the ``weight'' of fermionic pairs, and corresponds to the coherence  factor in the BCS state. At $r\rightarrow 0$, the SUSY Laughlin wavefunction is reduced to the original Laughlin wavefunction, while at $r\rightarrow \infty$, to the Moore-Read wavefunction.  
Interestingly, the SUSY pseudo-potential Hamiltonian is non-Hermitian and parameter-dependent, but its eigenvalues are real and parameter-independent. In this sense, the SUSY pseudo-potential Hamiltonian appears to realize an example of Bender's non-Hermitian Hamiltonian \cite{benderRPP2007}. We also discussed close relations between SUSY QH state and BCS superconducting state. 
%An analogous parameter-dependent SUSY wavefunction plays a crucial role in %the study of  doped valence bond antiferromagnetic models, and detailed %discussions are reported in Ref.\cite{SUSYAKLTpaper}.

%%%%%%%%%%%%%%%%%%%%%%%%%%%%%%%%%%%%%%%%%%%%%%%
%%%%%%%%%%%%%%%%%%%%%
%\begin{table} 
%\renewcommand{\arraystretch}{1}
%\begin{center}
% \begin{tabular}{|c|c|c|}  
% \hline    & The original (bosonic) set-up & ~~~~~~~~~~ The SUSY set-up ~~~~~~~~~  
%\\
% \hline 
%Base manifold & $S^{2}\!\simeq \!SU(2)/U(1)$ &  $S^{2|2}\!\simeq\! UOSp(1|2)/U(1)$
%\\ \hline  
%Monopole & Dirac monopole  &   Supermonopole
%\\
%\hline    
%  Hopf map & $S^3 \rightarrow S^2$ & $S^{3|2} \rightarrow5%S^{2|2}$  
%\\ 
%\hline  
%Fuzzy manifold & Fuzzy sphere, hyperboloid & Fuzzy supersphere,  superhyperboloid
% \\
%\hline
%Groundstate  & $SU(2)$ invariant Laughlin & $UOSp(1|2)$ invariant Laughlin 
%\\
%\hline
% Hamiltonian  & Hermitian   & Non-Hermitian 
%\\
%\hline
%\end{tabular}
%\end{center}
%\caption{ The original bosonic set-up and the SUSY extension.} 
%\end{table}\label{SUSYextensionofHaldane}
%%%%%%%%%%%%%%%%%%%%%%%%%%%%%
%%%%%%%%%%%%%%%%%%%%%%%%%%%%%%%%%%%%%%%%%

%\section*{Acknowledgements}

This work was supported by both Inoue and Sumitomo Foundations.

\end{document}